\documentclass[USenglish,twocolumn]{article}

\ifx\directlua\undefined\ifx\XeTeXcharclass\undefined
  \usepackage[utf8]{inputenc}                           
  \else\RequirePackage[no-math]{fontspec}[2017/03/31]\fi 
  \else\RequirePackage[no-math]{fontspec}[2017/03/31]\fi 
\usepackage[sort&compress,square,numbers]{natbib}
\usepackage[big,online]{dgruyter}
\usepackage{dsfont}
\usepackage{mathtools}

\theoremstyle{dgthm}

\theoremstyle{dgdef}

\DeclarePairedDelimiter\bra{\langle}{\rvert}
\DeclarePairedDelimiter\ket{\lvert}{\rangle}
\newcommand{\braket}[2]{\langle #1 \vert #2 \rangle}
\newcommand{\abs}[1]{|#1|}

\begin{document}

\articletype{Research Article}

\author*[1]{Helgi Sigurðsson}
\author[2]{Hai Chau Nguyen}
\author[3]{Hai Son Nguyenr} 
\affil[1]{Institute of Experimental Physics, Faculty of Physics, University of Warsaw, ul.~Pasteura 5, PL-02-093 Warsaw, Poland; \& Science Institute, University of Iceland, Dunhagi 3, IS-107 Reykjavik, Iceland, helg@hi.is; https://orcid.org/0000-0002-4156-4414}
\affil[2]{Naturwissenschaftlich-Technische Fakult\"at, 
Universit\"at Siegen, Walter-Flex-Stra{\ss}e 3, 57068 Siegen, Germany, Chau.Nguyen@uni-siegen.de; https://orcid.org/0000-0001-7204-3454}
\affil[3]{Univ Lyon, Ecole Centrale de Lyon, INSA Lyon, Universit\'e  Claude Bernard Lyon 1, CPE Lyon, CNRS, INL, UMR5270, Ecully 69130, France, hai-son.nguyen@ec-lyon.fr; https://orcid.org/0000-0002-1638-2272}

\title{Dirac exciton-polariton condensates in photonic crystal gratings}

\abstract{Bound states in the continuum have recently been utilized in photonic crystal gratings to achieve strong coupling and ultralow threshold condensation of exciton-polariton quasiparticles with atypical Dirac-like features in their dispersion relation. Here, we develop the single- and many-body theory of these new effective relativistic polaritonic modes and describe their mean-field condensation dynamics facilitated by the interplay between protection from the radiative continuum and negative-mass optical trapping. Our theory accounts for tunable grating parameters giving full control over the diffractive coupling properties between guided polaritons and the radiative continuum, unexplored for polariton condensates. In particular, we discover stable cyclical condensate solutions mimicking a driven-dissipative analog of the {\it zitterbewegung} effect characterized by coherent superposition of ballistic and trapped polariton waves. We clarify important distinctions between the polariton nearfield and farfield explaining recent experiments on the emission characteristics of these long lived nonlinear Dirac polaritons.}
\journalname{Nanophotonics}
\journalyear{2023}

\maketitle

\section{Introduction}
Quantum fluids with both unusual dispersive properties and strong non-Hermitian effects form a new exciting testbed to investigate many-body physics. Macroscopic quantum fluids of exciton-polaritons~\cite{Carusotto_RMP2013} (hereafter, {\it polaritons}) are particularly suited for this task given their optical malleability~\cite{Tosi_NatPhys2012, Pickup_NatComm2020} and readout, high interaction strengths, and the wide breadth of materials permitting polariton condensation at elevated temperatures~\cite{Sanvitto_NatMat2016, Jiang_AdvMat2022}. These light-mass bosonic quasiparticles form in the strong coupling regime between confined photonic modes and exciton resonances in semiconductor microcavities~\cite{Carusotto_RMP2013}. In particular, there has been some excitement in simulating relativistic phenomena in artificial Dirac materials exploiting the polariton spin in both real and synthetic magnetic fields~\cite{Tercas_PRL2014, Gianfrate_Nature2020, Su_SciAdv2021, Polimeno_Optica2021, Lempicka_SciAdv2022, lovett2023observation} and patterned photonic structures~\cite{Jacqmin_PRL2014,  Kexin_PRB2019, Klembt_Nature2018, Milicevic_PRX2019, Liu_Science2020, Li_NatComm2021, Wang_NSR2022}. Such materials, hosting associated Dirac cones, offer valuable insight to a plethora of exotic phenomena such as quantum Hall physics~\cite{Gianfrate_Nature2020}, nontrivial topological phases~\cite{Klembt_Nature2018}, Weyl semimetals~\cite{Lu_Science2015}, and relativistic trapping~\cite{Lee_Nanopho2021, Chen_SciAdv2023} while supplemented with strong polariton nonlinearities. Moreover, alternative neighbouring platforms for exploration into light-matter Dirac physics involve phonon-polaritons~\cite{Schmidt_NJP2015, Guddala_Science2021} and plasmon-polaritons~\cite{In_LightSci2022}.

Recently, a photonic crystal platform was realized to explore Dirac physics using exciton-polaritons. It consists of a subwavelength grated GaAs-based semiconductor waveguide embedded with multiple quantum wells hosting Wannier-Mott excitons (see Fig.~\ref{fig1}). Both strong light-matter coupling and ultralow threshold polariton Bose-Einstein condensation into the waveguide's associated bound-states-in-the-continuum (BIC) was demonstrated in the same study~\cite{Ardizzone_Nature2022}. A parallel study demonstrated the impressive tunability such grated waveguides offered over the polariton properties~\cite{Hu_APL2022}. In these systems, strong coupling is facilitated by the photonic structure's protection from the continuum~\cite{Azzam_AdvOptMat2021, Hwang_CommPhys2022} which allows photons to survive long enough to form polariton states~\cite{Lu_PhotRes2020, Zanotti_PRB2022}. The initial experiment~\cite{Ardizzone_Nature2022} was soon followed with fascinating results on the behaviour of the fluid's elementary excitations~\cite{Grudinina_NatCommun2023} and demonstration of macroscopic hybridization between coupled BIC condensates~\cite{gianfrate2023optically}. 
Besides III-V semiconductor photonic crystals~\cite{Bajoni_PRB2009}, other platforms able to show BIC-facilitated strong exciton-photon coupling consist of dry transfer deposited transition metal dichalcogenide monolayers such as MoSe$_2$~\cite{Kravtsov_LSA2020, Koksal2021} or WS$_2$~\cite{Zhang_NatComm2018, Maggiolini_NatMat2023, Weber_NatMat2023}; or spin coated hybrid organic-inorganic perovskites~\cite{Dang2020,Dang_AdvOptMat2022,Kim2021,Wang2023}.

Inspired by these rapid developments in BIC facilitated polariton condensation, and the surging interest to simulate nonlinear and non-Hermitian relativistic physics in an optically addressable setting~\cite{Leykam_PRL2017}, we develop the two-band theory of BIC Dirac exciton-polaritons in photonic crystal gratings. We then propose a many-body description in the mean field picture, allowing us to construct a simplified generalized two-band Gross-Pitaevskii model describing the interplay between BIC facilitated condensation and negative-mass optical trapping of Dirac polaritons. Our theory is in excellent agreement with recent experiments on polariton crystal gratings under nonresonant excitation~\cite{Ardizzone_Nature2022, Riminucci_PRL2023, gianfrate2023optically} and uncovers fascinating cyclical condensate dynamics through continuous adjustment of experimentally accessible parameters.

In particular, our mean field simulations reveal that the condensate negative-mass population ``drops'' iteratively to higher order optical trap modes as a function of pump power density, underlining complex interplay between pump induced polariton energy shifts and gain. Around the drop point the condensate can converge into a limit cycle solution, characterized by a coherent superposition of two distinct trap levels. More interestingly, through careful tuning of the photonic grating the BIC can be gradually moved from the lower negative-mass branch to the upper positive-mass branch, until the condensate suddenly stabilizes into a {\it zitterbewegung}-like~\cite{Sedov_PRB2018, lovett2023observation} limit cycle, forming a strange mixture of confined low momentum negative-mass polaritons and ballistic high momentum positive-mass polaritons. Lastly, we elucidate on how the polariton field within the photonic crystal grating relates to the emitted far field measured in experiment, in sharp contrast to typical condensation experiments in planar microcavities~\cite{Carusotto_RMP2013}.

\section{Massive Dirac polaritons}
\subsection{Photonic modes in optical grating}

\begin{figure}
\centering
\includegraphics[width=0.99\linewidth]{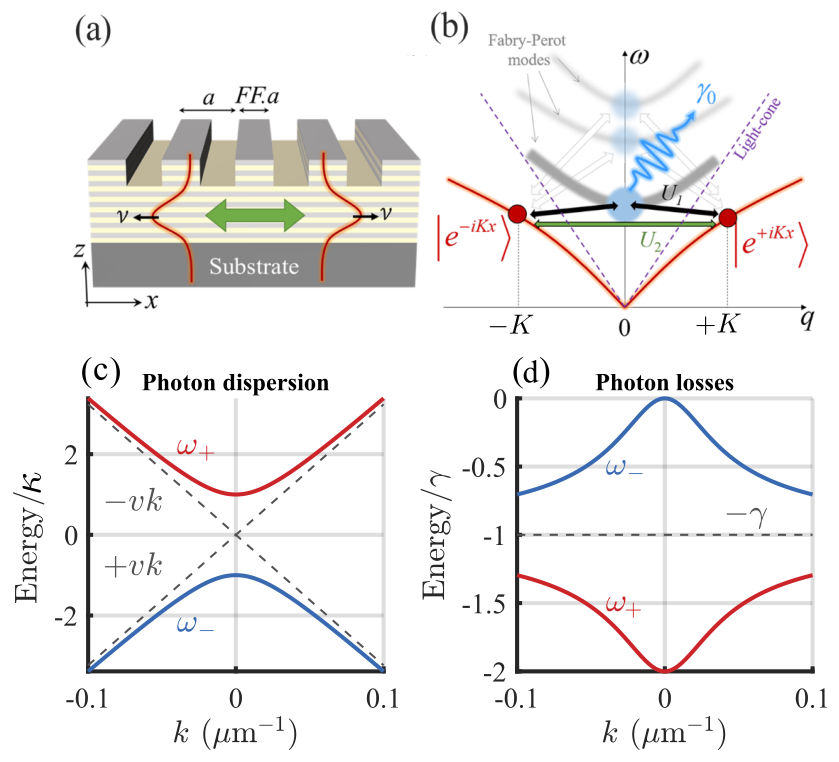}
\caption{\textbf{Scheme of the system and guided photonic modes.} (a) Schematic of the 1D subwavelength grated waveguide with period $a$ and filling fraction $FF$. The different colored layers denote e.g. GaAs quantum wells separated by AlGaAs barriers. (b) Sketch of the two guided modes $\ket{e^{\pm iKx}}$ coupled with each other and coupled to the Fabry-Pérot lossy modes through the grating. (c,d) Dashed black lines show the real and imaginary parts of the dispersion relation of uncoupled counter-propagating photons. Blue and red solid curves correspond to finite diffractive coupling leading to gap opening and formation of symmetric and antisymmetric standing wave modes, respectively. The BIC can be seen from the absence of losses in the antisymmetric mode at $k=0$. We have used: $\varphi=0$, $\hbar v = 56.6$ meV $\mu$m, $\hbar \kappa  = 1.75$ meV, and $\hbar\gamma  = 0.18$ meV corresponding to physical values obtained from rigorous coupled-wave analysis (see Methods).}
\label{fig1}
\end{figure}

We consider a waveguide stack consisting of multiple periodic layers of GaAs quantum wells separated by AlGaAs barriers along the normal $z$-direction [see Fig.~\ref{fig1}(a)]. On its upper surface, the waveguide is patterned with a one-dimensional subwavelength grating with a period $a$ along the $x$-direction and filling factor $FF$, forming a one-dimensional (1D) photonic crystal slab. This kind of a sample has been demonstrated in~\cite{Ardizzone_Nature2022, Hu_APL2022, gianfrate2023optically}, however our developed theory is also suitable for photonic crystals deposited with 2D materials~\cite{Low_NatMat2017,Zhang_NatComm2018,Kravtsov_LSA2020,Koksal2021, Maggiolini_NatMat2023, Weber_NatMat2023} or perovskites~\cite{ Dang2020,Dang_AdvOptMat2022,Kim2021,Wang2023}.

From here on we will consider only propagation of electromagnetic modes along the $x$-direction.
We concentrate also on the TE mode, where the polarization of the electromagnetic field is assumed to be along the $y$-axis. In the absence of the grating, the guided electromagnetic modes are plane waves $\ket{e^{iqx}}$ with momentum $q$ and frequency $\omega_q$. As is well known~\cite{joannopoulos2011photonic}, the grating with period $a$ can be modeled by a periodic potential acting on the photonic modes,
\begin{equation} 
V_{\mathrm{ph}}(x,z)=u(x) w(z)
\end{equation}
where $u(x)$ is a periodic function of period $a$, and $w(z)$ is a square-step function that is equal to $1$ for $z$ within the patterned layers and vanishes otherwise.

The periodic potential couples photonic modes with wavevectors different by integer number of the primitive reciprocal lattice number $K = 2 \pi/a$, a condition known as {\it Bragg reflection} which folds photonic bands over the first Brillouin zone. Here the period $a$ is chosen such that the exciton energy is around the frequencies $\omega_{\pm K}$. Therefore, the relevant photonic modes for exciton-photon coupling correspond to those with wavevectors $q = \pm K + k$ with $k \ll K$. In this range of frequencies, the periodic potential couples nearly degenerate guided modes $\ket{e^{i(K+k)x}}$ and $\ket{e^{i(-K+k)x}}$, giving rise to an effective Dirac Hamiltonian describing the dynamics of the electromagnetic waves. 
Remarkably, the periodic potential not only couples the two modes $\ket{e^{\pm iKx}}$ together, but also couples them with lossy modes residing at normal incidence $q=0$. This  renders the two guided modes $\ket{e^{\pm iKx}}$ eventually lossy; see Fig.~\ref{fig1}(b). In total, the dynamics of the electromagnetic waves in the photonic crystal grating can be described by a lossy Dirac Hamiltonian,
\begin{equation} \label{eq.Hpho}
H_{\text{ph}}=  \begin{pmatrix}
 v k & \kappa \\
\kappa & - v k
\end{pmatrix} -i\gamma\begin{pmatrix}
 1 & e^{i\varphi} \\
e^{-i\varphi} & 1
\end{pmatrix},
\end{equation}
with 
\begin{equation} \label{eq.parameters}
\kappa =|U_2|, \quad
\varphi =2 \operatorname{arg} (U_1) - \operatorname{arg}(U_2), \quad
\gamma \propto |U_1|^2,
\end{equation} 
where $U_p$ is $p$th Fourier coefficient of $u(x)$, i.e., $U_p=\bra{1} u(x) \ket{e^{-ipKx}}$.
We have also linearised the dispersal relation around $K$ so that $\omega_{k\pm K} = \omega_K \pm vk$ and ignore the constant term. A more detailed derivation of this effective low-momentum photonic Hamiltonian is given in the Methods.

Equation~\eqref{eq.Hpho} is a non-Hermitian Dirac Hamiltonian of which the coupling between counter-propagating guided photon modes depends on the waveguide diffraction mechanism of strength $\kappa$ and the loss exchange mechanism of strength $\gamma e^{i\varphi}$ via the radiative continuum~\cite{Lu_PhotRes2020, footnote1}. 
Notice that we adopt here the convention of using $e^{-i\omega t}$ to describe the temporal oscillation, hence losses are given by the negative imaginary component of the dispersion relation. 

Following Eq.~\eqref{eq.parameters}, the parameters $\kappa$, $\gamma$ and $\varphi$ are  dictated by the two first Fourier components $U_1$ and $U_2$ of the periodic modulation $u(x)$. The diffractive coupling $\kappa>\gamma$ is the main parameter responsible for the bandgap opening at $k=0$ [see solid curves in Fig.~\ref{fig1}(c)]. Its value can be engineered by tuning the filling fraction of the grating~\cite{Lu_PhotRes2020}; see Methods for numerical values of $\kappa$ in realistic sample designs. 
  
The dispersion relation of the Hamiltonian~\eqref{eq.Hpho}, corresponding to the new symmetric ($+$) and antisymmetric ($-$) standing-wave eigenmodes, can be found as
\begin{equation} \label{eq.pho}
\omega_{\pm} =  - i \gamma \pm \sqrt{(vk)^2 + \kappa^2 - \gamma^2 -2 i\gamma\kappa \cos{(\varphi)}}.
\end{equation}
In the absence of losses one recovers the Dirac dispersal relation with the effective photon Dirac mass $= \kappa$. The real and imaginary parts of the photon frequencies are plotted in Fig.~\ref{fig1}(c,d) for $\varphi = 0$ showing a zero-loss BIC at $k=0$ in the antisymmetric branch. This can be understood from,
\begin{align} \notag
\text{Im}{(\omega_{\pm})} & = -\gamma \mp  \sin{\left[ \frac{1}{2} \tan^{-1}{ \left( \frac{2 \kappa \gamma \cos{(\varphi)}}{(vk)^2 + \kappa^2 - \gamma^2} \right)} \right]} \\
& \times \sqrt[4]{\left[(vk)^2 + \kappa^2-\gamma^2\right]^2 + 4 \kappa^2 \gamma^2 \cos^2{(\varphi)}}.
\end{align}
For small loss, $\gamma \ll \kappa$, and $k\to0$, we have
    \begin{equation}\label{eq:gamma_pm}
    \lim_{k\to0}  \text{Im}{(\omega_{\pm})} \approx -\gamma(1 \pm \cos{\varphi}).
    \end{equation}
If the grating design possesses mirror symmetry, $x\rightarrow -x$, we have $u(x)=u(-x)$ and all Fourier coefficients of $u(x)$ are real. Hence, $\varphi$ can only take values that are integer multiple of $\pi$. Notably, a $\pi$-jump of $\varphi$ can be obtained by sweeping the filling fraction through a band-inversion point; see Methods. From \eqref{eq:gamma_pm}, in the case of $\varphi = 0$, a BIC mode of infinite lifetime appears in the center of the lower $\omega_-$ branch while a lossy $2 \gamma$ mode appears in the upper $\omega_+$ branch [see Fig.~\ref{fig1}(d)]. 
    \begin{equation}
    \lim_{k\to0}  \text{Im}{(\omega_{\pm})}\Big|_{\varphi = 0} = \begin{cases}
      -2\gamma \\    
      0    
    \end{cases}.
    \end{equation}
Conversely, when $\varphi=\pi$ the BIC switches branches. In both case, as long as $\varphi$ is an integer multiple of $\pi$ due to the presence of the mirror symmetry $x\rightarrow -x$, a formation of BIC at $\Gamma$ point is guaranteed. This BIC is therefore of a symmetry-protected nature~\cite{Azzam_AdvOptMat2021}.

 Breaking the mirror symmetry will relax the  aforementioned  constraint on $\varphi$. As a result, none of the branches exhibit a BIC. If the symmetry breaking is only a small perturbation, the symmetry-protected BIC becomes a quasi-BIC with finite but extremely long lifetime~\cite{quasiBIC1,quasiBIC2,quasiBIC3,quasiBIC4,quasiBIC5}.  We note that a similar form of~\eqref{eq.Hpho} has been previously reported to describe the formation of symmetry-protected BICs~\cite{Lu_PhotRes2020,Ardizzone_Nature2022}, however our work provides the first effective Hamiltonian for the general case where the in-plane mirror-symmetry can be broken. Importantly, for realistic grating structures with lateral symmetry breaking design, a fine tuning of $\varphi$ from 0 to $\pi/2$ can be achieved (see Methods).

 To start with, we will explore the case of a grating with lateral mirror symmetry that has $\varphi = 0$. At a later stage we will relax this constraint and explore Dirac-polariton BIC condensation when $0<\varphi<\pi$.

\subsection{Coupling between photonic BIC and excitons}
 We now consider the strong light-matter coupling regime between the photons and quantum well excitons leading to new hybrid modes known as exciton-polaritons~\cite{Carusotto_RMP2013}. Our goal is to design a simple mean field model describing the dynamics of a driven Bose-Einstein condensate of Dirac-polaritons. We start in the single particle limit where a standard coupled oscillator model can be written to describe the mixing of standing-wave photons and excitons with a light-matter coupling parameter $\Omega$ (also known as the exciton-photon Rabi frequency)~\cite{Gerace_PRB2007, Lu_PhotRes2020}. In our photonic grating, the lossy mode and BIC mode exhibit opposite parities of mirror symmetry $x\rightarrow -x$, with one being symmetric and the other being antisymmetric. Their near-field distributions are periodic functions with a period of the grating ($a\approx300$ nm), with the anti-nodes of the lossy mode field being the nodes of the BIC field, and vice versa. Therefore, excitons, tightly localized within the Bohr radius of about 10 nm, can only couple efficiently to either the lossy mode or the BIC mode due to the spatial overlap requirement. We then follow the approximation proposed by Lu et al~\cite{Lu_PhotRes2020} to divide excitons into two groups: one that only couples to the lossy mode and another that only couples to the BIC mode with the same coupling strength. Therefore, the strong coupling picture is dictated by a 4$\times$4 Hamiltonian, given by~\cite{Lu_PhotRes2020}:
\begin{equation} \label{eq.HLM}
H_{\text{pl}} = 
\begin{pmatrix} 
\omega_{+} & \Omega & 0 & 0 \\
\Omega &  \omega_X & 0 & 0 \\
0 & 0 & \omega_{-}  & \Omega  \\
0 & 0 & \Omega & \omega_X  \\
\end{pmatrix}.
\end{equation}
Here, $\omega_X = \omega^{(0)}_X - i \gamma_\text{nr}$ where $ \omega^{(0)}_X$ and $\gamma_\text{nr}^{-1}$ denote the detuning of the excitons from the photon branches at $k=0$ and their nonradiative lifetime, respectively. Here we have assumed that the mass of excitons is practically infinite compared to the confined photons. We note that there is no direct coupling from the excitons to the radiative continuum, only to the localized waveguided modes.  We highlight that this simple model successfully matches the numerical simulation results of polaritonic branches (see Section \ref{sec:simulVStheory}). In a more precise consideration, one should include a third group of excitons that are not localized at the antinodes of both photonic modes, thus do not undergo the strong coupling regime. This corresponds to the observation of uncoupled excitons shown in the simulations of Section 6.4. Since these uncoupled excitons do not affect the strong coupling physics, we do not include them in our effective theory.

\begin{figure}[t]
\centering
\includegraphics[width=0.99\linewidth]{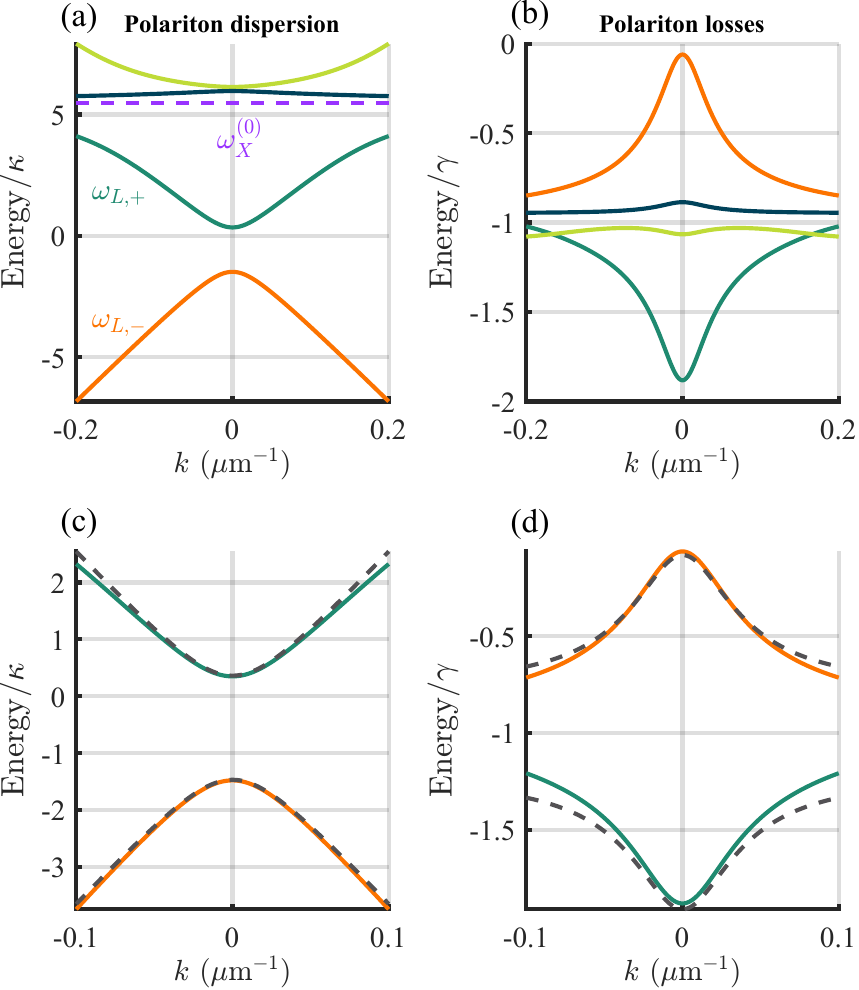}
\caption{\textbf{Dirac exciton-polariton dispersion relation.} Solid lines show the (a) real and (b) imaginary energies of the polariton dispersion from Eq.~\eqref{eq.pol} with the exciton line and the lower symmetric $\omega_{L,+}$ and antisymmetric $\omega_{L,-}$ branches marked. (c,d) Show a zoom of the real and imaginary energies belonging to the lower symmetric and antisymmetric polariton states. The dashed black lines show the approximation obtained using~\eqref{eq.heur}. Here we set: $\varphi=0$, $\Omega/\kappa= 1.8$, $\hbar\omega_X^{(0)}/\kappa = 5.5$, and $\gamma_\text{nr} = \gamma$ corresponding to physical values obtained from RCWA (see Methods).}
\label{fig2}
\end{figure}

The eigenmodes of the Hamiltonian \eqref{eq.HLM} are referred to as upper $|U,\pm\rangle$ and lower $|L,\pm\rangle$ symmetric-antisymmetric polaritons with a dispersion relation,
\begin{align} \label{eq.pol}
\begin{split}
\omega_{U,\pm} &= \frac{\omega_{\pm} + \omega_X}{2} + \frac{1}{2} \sqrt{ (\omega_{\pm} -\omega_X)^2 + 4\Omega^2},\\
\omega_{L,\pm} &= \frac{\omega_{\pm} + \omega_X}{2} - \frac{1}{2} \sqrt{ (\omega_{\pm} -\omega_X)^2 + 4\Omega^2},
\end{split}
\end{align}
which is plotted in Fig.~\ref{fig2}(a) and~\ref{fig2}(b). Notice that due to the exciton losses, the BIC now becomes a quasi-BIC with finite losses.

We will assume that the upper polariton branches $\omega_{U,\pm}$ are far away in energy and weakly populated and thus only focus on the lower polariton branches $\omega_{L,\pm}$ around $k=0$ where condensation preferentially takes place~\cite{Ardizzone_Nature2022, Riminucci_PRAppl2022, gianfrate2023optically}. The lower branches can be approximated by considering first the coupling of forward- $\ket{e^{ iKx}}$ and backward-propagating $\ket{e^{-iKx}}$ photons to excitons, leading to lower-forward $\ket{L,e^{iKx}}$ and lower-backward  $\ket{L,e^{-iKx}}$ propagating polaritons. This is followed up by the photonic diffractive coupling mechanism evaluated at small momenta. In other words, polariton modes are first calculated in the grating free waveguide then we perturbatively introduce the grating back (diffractive coupling). The lower polariton dispersion can then be written,
\begin{equation} \label{eq.heur}
\omega_{L,\pm} \approx  \tilde{\omega}_L \pm \sqrt{(\tilde{v}k)^2 +  [\kappa^2 - \gamma^2 -2 i\gamma\kappa \cos{(\varphi)}] |C_0|^4},
\end{equation}
where the first term in~\eqref{eq.heur} corresponds simply to an overall complex energy shift due to the light-matter coupling which is written,
\begin{equation} 
    \tilde{\omega}_L = \frac{\omega_X}{2}\left(1 -\sqrt{1+ \dfrac{4\Omega^2}{ \omega_X^2}}\right) -  \frac{i\gamma}{2} \left(1 + \frac{1}{\sqrt{1 + \dfrac{4 \Omega^2}{\omega_X^2}}}\right).
\end{equation}
The renormalized light-matter velocity $\tilde{v}$ and the photon Hopfield coefficient of forward and backward propagating lower polaritons around $k=0$ are given by,
\begin{align}
 \tilde{v} & = \frac{v}{2} \left(1  +  \frac{  \omega_X}{ \sqrt{ \omega_X^2 + 4\Omega^2}} \right),\\ \label{eq.hopf}
|C_0|^2 & = \frac{4\Omega^2}{4\Omega^2 +  \left|\omega_X +i\gamma - \sqrt{(\omega_X+i\gamma)^2 + 4\Omega^2}\right|^2 }. 
\end{align}

The form of Eq.~\eqref{eq.heur} implies that, in the truncated basis of lower forward $\ket{L,e^{iKx}}$ and backward $\ket{L,e^{-iKx}}$ polaritons, we can describe the system with the following massive non-Hermitian Dirac operator,
\begin{equation} \label{eq.HpolDirac}
H_D =  \tilde{\omega}_L \mathds{1}_{2\times2}  + \begin{pmatrix}
 \tilde{v} k & (\kappa - i \gamma e^{i \varphi}) |C_0|^2 \\
(\kappa - i \gamma e^{-i \varphi})|C_0|^2  & - \tilde{v} k
\end{pmatrix},
\end{equation}
with new symmetric and antisymmetric lower polariton eigenstates $|L,\pm\rangle = A_\pm \ket{L,e^{iKx}} + B_\pm \ket{L,e^{-iKx}}$ with eigenenergies $\omega_{L,\pm}$. One limitation of Eq.~\eqref{eq.heur} is that it neglects the dependence of the photonic Hopfield coefficient~\eqref{eq.hopf} on both momentum and the original diffractive coupling between the counterpropagating photons. However, if $\kappa \ll \Omega \ll \omega_X^{(0)}$ then Eq.~\eqref{eq.heur} remains accurate and implies that waveguided TE polaritons behave approximately as Dirac particles with renormalized velocity $\tilde{v}$ and gap opening. In Fig.~\ref{fig2}(c) and~\ref{fig2}(d) we compare our approximated dispersion~\eqref{eq.heur} (dashed lines) with the lower polariton dispersion relation coming from~\eqref{eq.pol} (solid lines) for values extracted from RCWA simulations, $\Omega/\kappa= 1.8$, $\hbar\omega_X^{(0)}/\kappa = 5.5$, and $\gamma_\text{nr} = \gamma$, and observe very good agreement. We note that the parameters used in our study accurately represent a real example of a photonic grating analyzed using RCWA (see Methods). The above underpins the feasibility in creating photonic samples that permit study of Dirac polariton quasi-BIC physics.

\section{Mean-field formalism}
\begin{figure*}[t]
\centering
\includegraphics[width=0.95\linewidth]{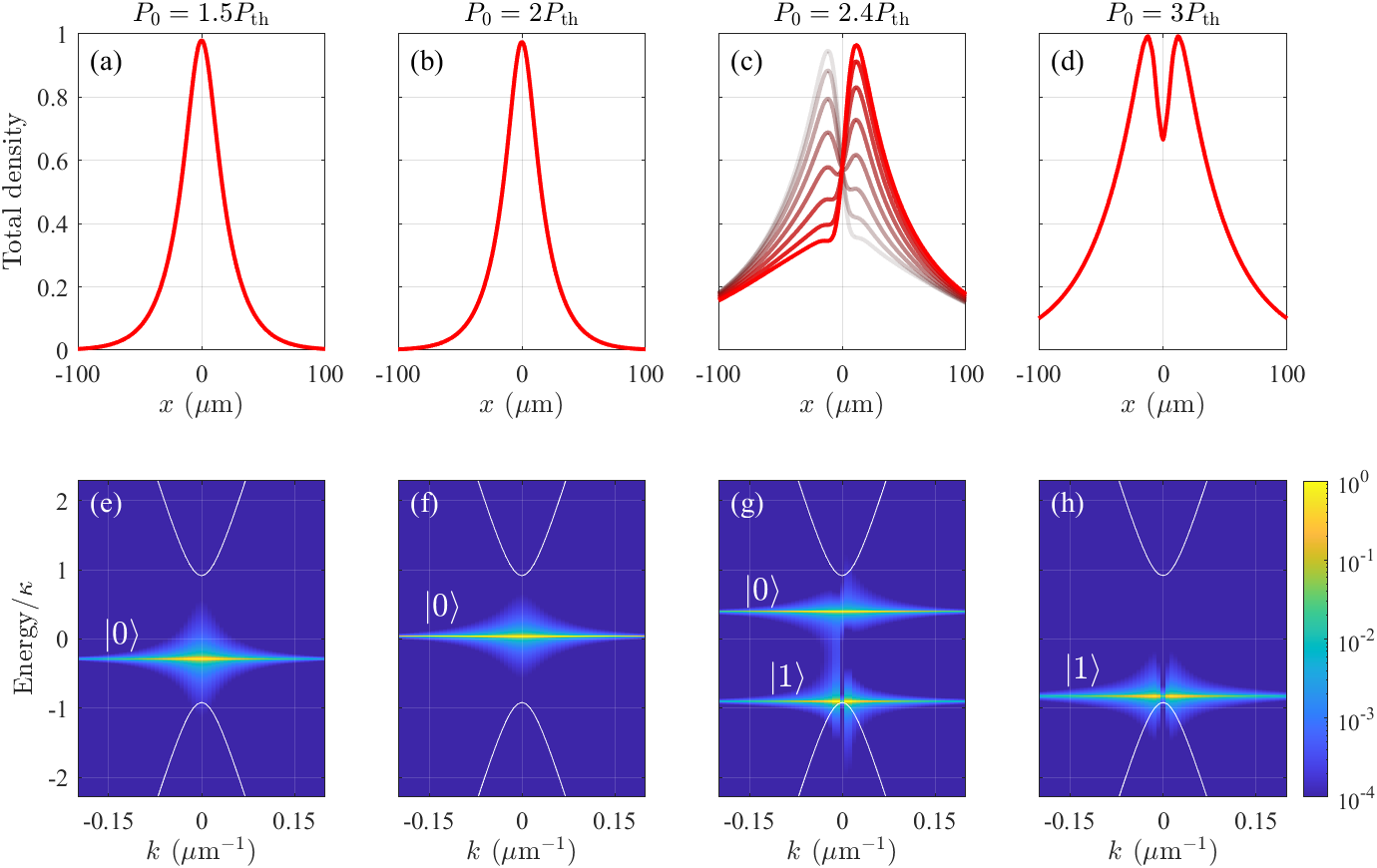}
\caption{\textbf{Mode dropping dynamics in simulated Dirac polariton BIC condensates.} Normalized total condensate density in (a-d) real space and (e-h) energy resolved momentum space for increasing pump power $P_0 = \{1.5,   2,   2.4,   3\} P_{\text{th}}$. The overlaid curves in panel (c) are taken at different time steps to show the nonstationary oscillations. Same parameters as in Fig.~\ref{fig2} are used here with a pump spot size of $\text{FWHM}=20$ $\mu$m (full width at half maximum). Note that we have shifted the zero energy point into the center of the bandgap. The labels $|0\rangle$ and $|1\rangle$ denote the condensate fraction occupying the trap ground state and first excited state, respectively. White solid curves denote the single-particle guided polariton dispersions in the photonic crystal from Eq.~\eqref{eq.heur}.}
\label{fig3}
\end{figure*}

To create a macroscopic coherent quantum state of the polaritons by means of Bose--Einstein condensation, the system is excited by an external nonresonant laser~\cite{Ardizzone_Nature2022, Riminucci_PRAppl2022}. This creates hot free charge carriers which relax in energy to form a reservoir of excitons at the so-called ``bottleneck region'' in the lower polariton dispersion relation~\cite{Wouters_PRL2007} denoted by the density parameter $n_R$. When then density of the pumped reservoir is sufficiently high, the polariton occupation number accumulated at a particular level (such as the quasi-BIC) can exceed unity and stimulated scattering of polaritons into this state starts. This signifies the spontaneous breaking of symmetry and non-equilibrium Bose-Einstein condensation into a single quantum state~\cite{Carusotto_RMP2013}, marked by a threshold power $P_\text{th}$. 
Following the mean-field theory~\cite{Wouters_PRL2007, Nigro_PRB2023} and our approximate single particle Dirac operator~\eqref{eq.HpolDirac}, the condensate can be described by a two-component macroscopic spinor wave function, or an {\it order parameter}, $\Psi(x,t) = (\psi_{+1}, \psi_{-1})^\text{T}$ where $\pm 1$ denotes the lower forward- and backward propagating polaritons, respectively. Note that subscripts $(\pm)$ without "1" denote the symmetric-antisymmetric basis which should not be confused here.

The generalized Gross-Pitaevskii equation for the condensate in real space can be found to be
\begin{align} \notag
   i \frac{\partial \Psi}{\partial t} & = \bigg[ H_D
   + g \Psi^{\dagger} \Psi
   \\ \label{eq:GP1}
   & +g_R \left(n_R + \frac{\eta P(x)}{\Gamma_R} \right) +i\frac{ R n_R}{2}
   \bigg]
   \Psi, \\ \label{eq:GP2}
       \frac{\partial n_R}{\partial t}  & = P(x) - (\Gamma_R +R \Psi^{\dagger} \Psi) n_R. 
\end{align}
The first term corresponds to the single particle dynamics~\eqref{eq.HpolDirac} in real space obtained by substituting $k \to -i \partial_x$. In the second term the polaritons are assumed to interact via short-range interaction described phenomenologically by the repulsive non-linear term $g \Psi^{\dagger} \Psi$. Moreover, polaritons also interact repulsively with strength $g_R>g$ against any background excitons whose density can be divided in to the bottleneck part $n_R$ and a static inactive {\it dark} exciton background parametrized by the dimensionless number $\eta$. The last term describes the stimulated scattering of reservoir excitons into the condensate at a rate $R$. The term $P(x) = P_0 e^{-x^2/2w^2}$ describes the continuous wave nonresonant Gaussian pump of waist $w$, and $\Gamma_R$ is the average reservoir exciton redistribution and decay rate.

For large waist $w$, the pumping can be considered to be uniform, $P(x) = P_0$. In this case, the condensation threshold is given by $P_\text{th} = - 2 \Gamma_R \text{max}\{\text{Im}{[\omega_{L,\pm}]}\}/R$ where $\text{Im}{[\omega_{L,\pm}]}<0$ and the maximum is taken over $\pm$. In this uniform case the maximum (i.e., minimum losses) will always correspond to the quasi-BIC. For a finite size pump the threshold actually increases due to finite gain region effects and must be determined numerically. For details on numerical modeling of the mean field equations please see Methods. 

Below threshold one has $\Psi^\dagger \Psi = 0$ in the long time limit and the reservoir converges to the steady state $n_R = P(x)/\Gamma_R$. We can then define a pump induced potential term acting on the Dirac polaritons,
\begin{equation} \label{eq.pot}
    V(x) = g_R(1 + \eta) \frac{P(x)}{\Gamma_R} > 0.
\end{equation} 
The above expression gives a good estimate for the optical trap felt by the condensate when pumped only weakly above threshold. For the positive branch polaritons, $V(x)$ acts as a repulsive gain region which, if tightly focused into a small enough spot, results in so-called ballistic condensation~\cite{Pickup_NatComm2020,  Alyatkin_NatComm2021}. For the negative branch polaritons however, it acts as an attractive gain region, pulling in generated polaritons and trapping them efficiently with a much lower threshold~\cite{Ardizzone_Nature2022, gianfrate2023optically, Nigro_PRB2023}.

\subsection{Mode dropping}
Figure~\ref{fig3} shows the total condensate density $\rho = \Psi^{\dagger} \Psi$ in (upper row) real space and (lower row) energy-resolved momentum space for increasing pump power density $P_0$. At low powers above threshold [Fig.~\ref{fig3}(a,e)], the condensate first occupies the ground state in the effective optical trap since it is closest to the quasi-BIC and has the lowest particle losses~\cite{Nigro_PRB2023}. 

Increasing the power we observe monotonic blueshift of the condensate level [compare Fig.~\ref{fig3}(a) and~\ref{fig3}(b)]. By increasing the power, more trap states become available for the condensate to populate and we can locate stable cyclical solutions (i.e., limit cycles) in which the condensate becomes nonstationary and coherently divided between two neighbouring trap modes~\cite{Topfer_PRB2020} in the same branch. Such cyclical solutions [see Fig.~\ref{fig3}(c)] usually appear through Hopf bifurcations when one fixed point attractor deteriorates and another takes over as parameters of the system are tuned. When the power is further increased, the blueshift is so strong that the fundamental trap mode is swept into the upper positive-mass band with increased losses. Consequently, the condensate abandons the fundamental mode and shifts its population into the neighbouring higher-order mode at at lower energies [see Fig.~\ref{fig3}(d)]. In this sense, the condensate "drops" from one trap mode to the next, as predicted by Nigro {\it et al.}~\cite{Nigro_PRB2023}. Increasing the power further, we observe periodically the same mode-dropping behaviour as subsequent higher order trap modes form in vicinity of the quasi-BIC and blueshift up into the lossy positive-mass band.

This power driven change in the condensate structure is in agreement with recent experimental observations~\cite{Ardizzone_Nature2022}. Energetically, this behaviour is in sharp contrast to optically trapped polariton condensates in planar cavities~\cite{Sun_PRB2018, Topfer_PRB2020, Alyatkin_NatComm2021} where stronger pumping results in a condensate dropping into {\it lower order} trap modes until it reaches the ground state.

\subsection{Negative-positive mass superposition}
Next, we characterize the interplay between the quasi-BIC state and the negative-mass trapping mechanism coming from the localized pumping area [see Eq.~\eqref{eq.pot}]. As mentioned around Eq.~\eqref{eq.Hpho} the photonic grating introduces a complex coupling parameter between the counterpropagating photons (which carries into the polariton modes). Up until now, we have taken $\varphi = 0$ which leads to a quasi-BIC in the lower (symmetric) branch in Fig.~\ref{fig2}(c) and~\ref{fig2}(d). If $\varphi = \pi$ then the quasi-BIC would instead form in the upper branch~\cite{Lu_PhotRes2020}. We perform a scan across both $\varphi$ and the pump power $P_0$ and investigate the difference between symmetric and antisymmetric condensate occupation
\begin{equation}
\Delta_\rho = \rho_+ - \rho_-.
\end{equation}
where
\begin{equation} \label{eq.diffrho}
    \rho_\pm = |\langle \Psi_\pm | \Psi \rangle |^2.
\end{equation}
and $|\Psi_\pm \rangle$ are the symmetric and antisymmetric eigenstates of the potential-free Dirac operator~\eqref{eq.HpolDirac}. In this sense, the quantity $\Delta_\rho$ is similar to the projection of a two level quantum system onto the axis connecting the north and south antipodal points of the Bloch sphere. If $\Delta_\rho<0$ then most of the condensate forms in the lower branch, whereas if $\Delta_\rho > 0$ then the condensate forms in the upper branch. 
\begin{figure}[t]
\centering
\includegraphics[width=0.99\linewidth]{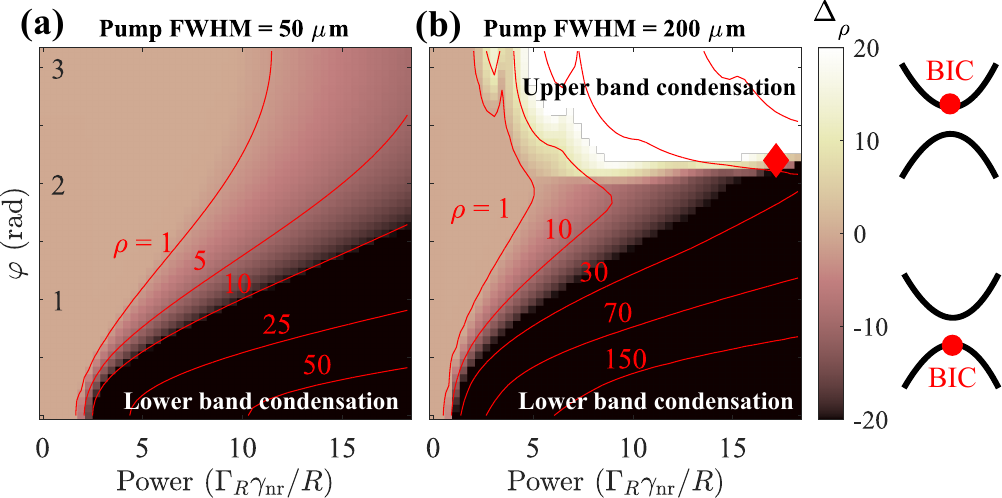}
\caption{\textbf{Condensation phase map as a function of BIC location and power.} Average condensate population difference $\langle \Delta_\rho \rangle$ between the symmetric and antisymmetric projections (branches) as a function of pump power, and the dissipative coupling parameter $\varphi$, and two different sizes of the pump spot. The colorscale is saturated around $\pm 1$ to more clearly show the white-dark regions. The red contours show total ($\rho = \Psi^\dagger \Psi$) isodensity curves. Other parameters are the same as in Fig.~\ref{fig3}.}
\label{fig4}
\end{figure}

The results are shown in Fig.~\ref{fig4} for two different sizes of pump spots (FWHM $ = 50$ and $200$ $\mu$m) where each pixel corresponds to the spatiotemporal average of $\langle \Delta_\rho \rangle$ over the entire simulation grid and integration time. The results show that for $\varphi \sim 0$, when the quasi-BIC is in the lower antisymmetric branch, we always have condensation in the same branch (dark region) as seen in experiments~\cite{Ardizzone_Nature2022}. Interestingly, for smaller pump spots in Fig.~\ref{fig4}(a) condensation still takes place in the lower antisymmetric branch even when the quasi-BIC has moved to the upper symmetric branch as can be seen from the weakly dark region at high powers around $\varphi \sim \pi$. This implies that the optical trapping mechanism can be more efficient in reducing transverse ($x$-direction) losses than the quasi-BIC in reducing out-of-plane ($z$-direction) losses. Nevertheless, the presence of the BIC has a dramatic effect on the condensation threshold curve approximately given by far-left red contour. Indeed, when the BIC is in the negative mass branch the optical trapping and protection from the continuum complement each other to lower the power needed to reach bosonic stimulation.

To explore the reduction of the optical trapping effect on lower branch polaritons we repeat the calculation using a much larger pump spot of FWHM $= 200$ $\mu$m in Fig.~\ref{fig4}(b). The much wider pump spot imposes a weaker confinement compared to the narrow spot. Indeed, now around $\varphi \sim \pi$ we see that condensation starts taking place in the upper branch (white region), following the quasi-BIC. This means that condensation can be optically adjusted between the lower and the upper branch by simply changing the size of the pump spot in a given sample. 
\begin{figure}[t]
\centering
\includegraphics[width=0.99\linewidth]{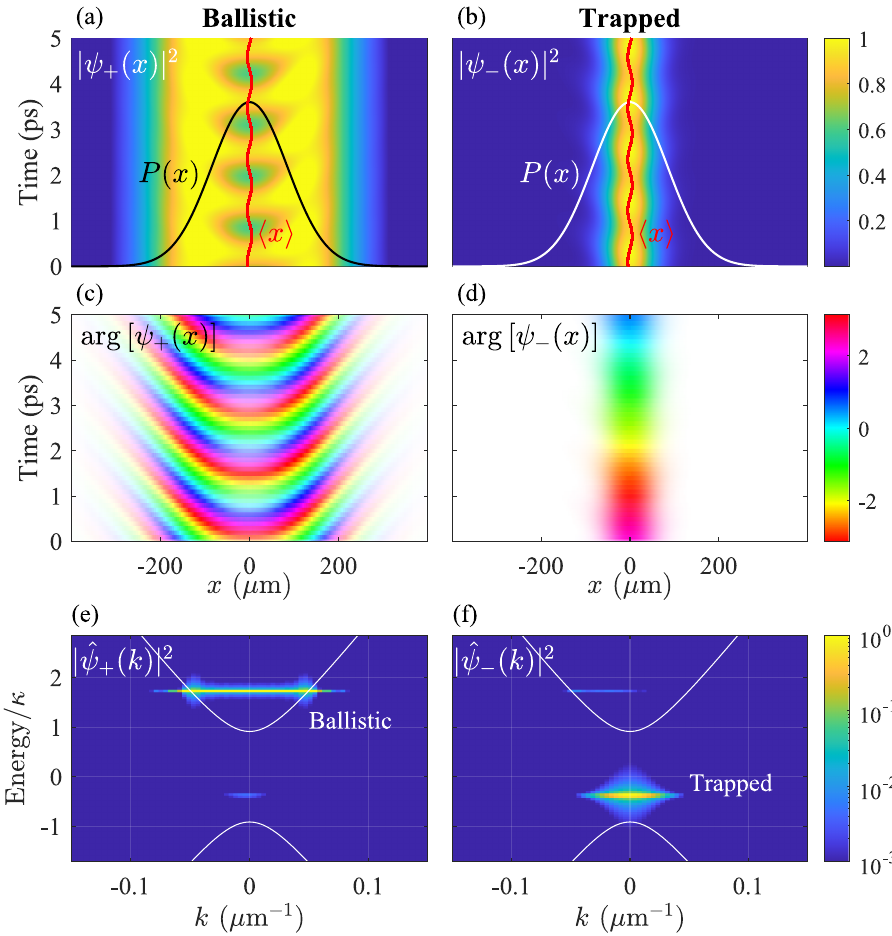}
\caption{\textbf{A stable superposition of trapped and ballistic polaritons with a single pump spot.} (a,b) Density and (c,d) phase of the symmetric and antisymmetric polaritons. The transparency of the phase map is proportional to the condensate density. Red trembling trajectory shows the center-of-mass of the condensate. (e,f) Corresponding condensate densities in Fourier space showing that each component belongs to different branches. Other parameters are the same as in Fig.~\ref{fig3}.}
\label{fig5}
\end{figure}

At the interface of such qualitatively different condensate solutions [i.e., dark and bright regions in Fig.~\ref{fig4}(b)] more exotic patterns might appear. We propose that by tuning the size of the trap, the grating pitch ($\varphi$), and pump power one can achieve simultaneous condensation in the upper and lower branches. This corresponds to a stable coherent mixture of positive and negative mass condensate polaritons (i.e., ballistic and trapped polaritons). Interestingly, such a solution bears similarities to the famous Zitterbewegung effect, the trembling motion of relativistic particles, but here in a driven-dissipative setting~\cite{Sedov_PRB2018}. Interference between the upper and lower branches causes the center-of-mass of the condensate, $\langle x \rangle = \int \Psi^\dagger \hat{x} \Psi dx / \int \Psi^\dagger \Psi dx$, to jitter in time (see red oscillating curve). In Fig.~\ref{fig5}(a,b) and (c,d) we show the density and phase of symmetric and antisymmetric condensate polaritons belonging to such a solution, obtained at the location of the red diamond marker in Fig.~\ref{fig4}(b). This solution is also a limit cycle, a stable coherent superposition of ballistic upper branch and trapped lower branch polaritons, with THz Rabi oscillations, $\propto \cos^2{(\pi \Delta E t/h)}$ (not to be confused with the light-matter Rabi frequency $\Omega$). Here $h$ is the Planck's constant and $\Delta E$ is the energy splitting between the ballistic and the trapped condensate levels [see Fig.~\ref{fig5}(e,f)]. The ballistic nature of the upper branch fluid manifests in its more delocalized nature and more rapidly varying phase profile along the $x$-direction.

\section{Nearfield and Farfield pattern}
In the previous sections, we demonstrated that the dynamics of polariton condensation is determined by the wavefunction and population of polaritons, derived from the generalized Gross-Pitaevskii equation. To probe polaritons in practical realizations, most experimental works rely on detecting their photonic component using far-field setups in either real or momentum space \cite{Ardizzone_Nature2022, gianfrate2023optically}. We remind the reader that the polariton state is explicitly related to the emitted photon state vector $\Psi \propto \Phi$ through the photonic Hopfield coefficient~\cite{Carusotto_RMP2013}. For polaritons in microcavities, the distinctions between nearfield and farfield, both directly determined by the polariton population, were often overlooked. 
Furthermore, recent studies show that nearfield setups can probe the local wavefunction of polaritons when they are not embedded in thick vertical microcavities~\cite{Mrejen2019}. 
As a result, it is crucial to bridge the polariton wavefunction with the pattern of the electric field in both nearfield and farfield scenarios. 
Interestingly, for a BIC, the relationship between nearfield and farfield turns out to be radically different. 

We note that there are certain inconsistent use of the terms `nearfield' and `farfield' in the literature of polaritonics. In this work, we adopt the conventional definition presented in Ref.~\cite{NearField_Farfield}: `nearfield' refers to the light field that is confined within structures and can only be probed using evanescent techniques such as scanning near-field optical microscopy (SNOM), while `farfield' pertains to the light field that propagates through space and can be probed using conventional imaging techniques.
This is different from corresponding notions that have been used in several polaritonic experiments~\cite{Amo2009,Anton2014,Estrecho2019,Pieczarka2020}, where `nearfield' actually refers to the usual farfield  measurements in real space, and `farfield' refers to the usual farfield measurements in momentum space. 

\begin{figure}
    \centering
    \includegraphics[width=0.99\linewidth]{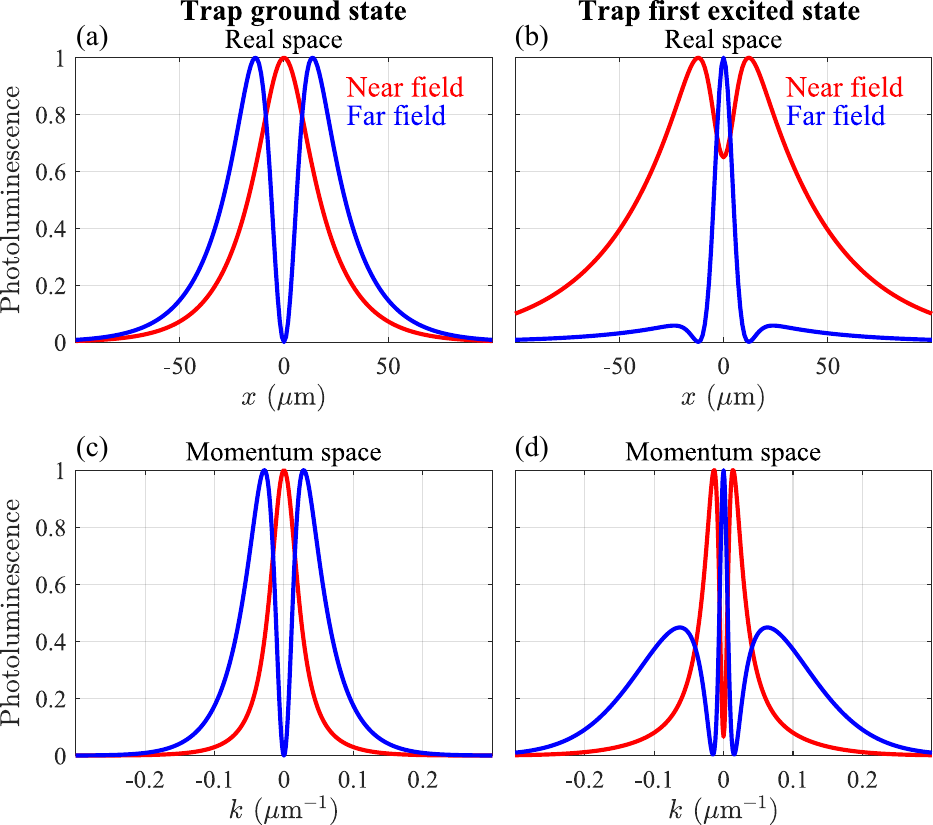}
    \caption{\textbf{Comparison of the nearfield and farfield polariton photoluminescence intensities.} Curves in all panels are individually normalized. (a,b) show the trap ground state and first excited state condensate emission corresponding to Figs.~\ref{fig3}(a) and~\ref{fig3}(d). (c,d) Show the corresponding momentum space emission belonging to Figs.~\ref{fig3}(e) and~\ref{fig3}(h).}
    \label{fig6}
\end{figure}

In fact, the nearfield and farfield pattern of BIC can be deduced rather  straightforwardly from the effective Dirac Hamiltonian for the photonic component~\eqref{eq.Hpho}. 
Indeed, we start with rewriting the Hamiltonian~\eqref{eq.Hpho} in real-space by substituting $k \to - i \partial_x$ and denoting the two-component photon state vector as $\Phi = (\phi_{+1}, \phi_{-1})^\text{T}$ where $\phi_{\pm1}$ are the coefficients of the forward and backward propagating photons $\ket{e^{i (k \pm  K) x}}$,
\begin{equation}
 H_\text{ph} =  \begin{pmatrix}
 -v i \partial_x & \kappa \\
\kappa & + v  i \partial_x
\end{pmatrix} 
-i \gamma
\begin{pmatrix}
 1 & e^{i\varphi} \\
e^{-i\varphi} & 1
\end{pmatrix}.
\end{equation}
Then using the dynamical equation $\partial_t \Phi (x,t) = - i H_\text{ph} \Phi(x,t)$ and its conjugation, we arrive directly at the continuity equation for the photon field,
\begin{equation}\label{eq:continuity}
    \partial_t (\Phi^{\dagger} \Phi) = -\partial_x (\Phi^{\dagger} \sigma_z \Phi) - \gamma \Phi^\dagger \begin{pmatrix}
 1 & e^{i\varphi} \\
e^{-i\varphi} & 1
\end{pmatrix} \Phi.
\end{equation}
where $\sigma_z$ is the third Pauli matrix. From the continuity equation, one infers, as usual, that $\Phi^{\dagger} \Phi$ describe the nearfield intensity, and $\Phi^{\dagger} \sigma_z \Phi$ describes the photon density current. The last term, 
\begin{equation}
    \Phi^\dagger \begin{pmatrix}
 1 & e^{i\varphi} \\
e^{-i\varphi} & 1
\end{pmatrix} \Phi = \gamma \abs{\phi_{+1} + e^{i \varphi} \phi_{-1}}^2,
\label{eq:farfield1}
\end{equation}
is then identified as losses by means of radiation into the farfield.
In this way, the expression for nearfield and farfield intensities have been obtained only by formally investigating the structure of the effective Dirac equation. Their difference can be appreciated from the additional interference term between the forward and backward propagating polaritons when~\eqref{eq:farfield1} is expanded. We show in the Methods section how they can also be understood from the microscopic theory. Figure~\ref{fig6} shows the difference between the nearfield (red curves) and farfield (blue curves) intensities of the emitted light from the condensate in both real space and momentum space corresponding to our results in Fig.~\ref{fig3}. Notably, in~\cite{Ardizzone_Nature2022} only the farfield was measured showing emission profiles which agree very well with our results. 

\section{Conclusions and discussion}
We have introduced the concept of Dirac polaritons in photonic crystal gratings---one dimensional photonic crystal slabs---containing excitonic resonances and symmetry protected photonic modes or {\it bound states in the continuum}. We developed the single particle theory of these effectively relativistic bosonic elementary excitations of light and matter, followed by intuitive extension to the many-body picture through the mean-field formalism. We propose a generalized Gross-Pitaevskii model to describe BIC-facilitated condensation of polaritons into pump-induced optical traps. Our findings are in excellent agreement with recent experimental observations on multi-quantum-well structures~\cite{Ardizzone_Nature2022} and are applicable to other forms of optically active materials such as transition metal dichalcogenide monolayers including MoSe$_2$~\cite{Kravtsov_LSA2020, Koksal2021} or WS$_2$~\cite{Zhang_NatComm2018, Maggiolini_NatMat2023, Weber_NatMat2023}; or hybrid organic-inorganic perovskites~\cite{Dang2020,Dang_AdvOptMat2022,Kim2021,Wang2023}.

Our theory is fully generalized towards photonic gratings with broken lateral symmetry which manifests in tunable diffractive coupling mechanism between guided photons and the continuum, allowing us to continuously tune the BIC from the lower energy Dirac branch to the upper. This gives powerful control over the polariton condensation threshold, and final state stimulation. We have also clarified on the distinction between farfield and nearfield emission patterns which becomes more important when dealing with subwavelength-grated photonic structures, and therefore will be important for future works on polaritonic crystals.

In particular, in mean-field simulations, we have identified peculiar zitterbewegung-like solutions in the driven Dirac polariton condensate which manifests in spontaneous formation of coherent superposition of upper-branch (positive) and lower-branch (negative) polaritons. The implications of such a hybrid quantum fluid are the two very different coupling mechanisms when neighbouring condensates are added. One one hand, the high energy component of the condensate will interact ballistically with its neighbours, on the other, the low energy component will interact evanescently with its neighbours. Such system could offer novel patterns of synchronicity between multi-component nonlinear oscillators with contrasting coupling mechanisms with competing coherence scales and time-scales of domain wall formation.

\section{Methods}
\subsection{Derivation of the non-hermitian Hamiltonian}\label{sec:derivationH}

Here we derive the effective photonic Dirac Hamiltonian~\eqref{eq.Hpho} from the microscopic consideration. To develop a perturbative theory for the guided photons, we first remark that in the absence of the modulated refraction index in the $x$-direction,  the eigenmodes of the lowest band of the system with frequency $\omega_q$ are the plane waves $\ket{e^{i q x}}$. For now, we ignore the confinement of the wave function in $z$-direction, which is assumed to only weakly dependent on $q$.
We also ignore the free evolution in the $y$-direction and assume that the system is time-reversal symmetric $\omega_q=\omega_{-q}$ without any loss of generality. Note that modes below the light cone are lossless, while the one above the light cone are lossy; see Fig.~\ref{fig1}(a).

The periodic modulation---with period $a$---of the refraction index in the $x$ direction introduces an periodic potential $u(x) = u(x+a)$ acting on the photons. As an expression of the Bragg reflection, the potential then couples modes $\ket{e^{i q x}}$ with that of $\ket{e^{i (q + nK) x}}$ where $K=2\pi/a$ is the primitive reciprocal lattice vector and $n$ is an integer. The relevant matrix elements are written $U_n= \bra{e^{i(q+nK)x}} u(x) \ket{e^{iqx}}$, which are the $n$-th Fourier coefficients of the potential. 
In particular, for any integer $n$, the potential couples degenerate modes of the same frequency, $q=nK/2$ and $q=-nK/2$.  

We are interested in the system excited at frequencies corresponding to $q$ around $\pm K$. 
The relevant wavevectors are therefore of $\pm (K+k)$ with $k \ll K$.
The two modes $\ket{e^{i(k+K)x}}$ and $\ket{e^{i(k-K)x}}$ are coupled by the potential $u(x)$ with matrix elements $U_2= \bra{e^{i(k+K)x}} u(x) \ket{e^{+i(k-K)x}} = \bra{1} u(x) \ket{e^{-i2Kx}}$ and $U_2^\ast= \bra{1} u(x) \ket{e^{+i2Kx}}$, which are simply the second Fourier coefficients of $u(x)$.

Being guided modes located below the light-line, both modes $\ket{e^{i(k\pm K)x}}$ are technically lossless. They become, however, lossy through coupling to lossy modes at low momentum, $\ket{e^{ikx}}$. Notice that these lossy modes at $\ket{e^{ikx}}$ are distributed on the Fabry-Pérot modes of the stack; see Fig.~\ref{fig1}(b). Therefore, in order to describe matrix elements of these scattering processes, we need to include the confinement wave function in the $z$-direction.
Including these confinement factors, the full wavefunctions of the two modes $\ket{e^{i(k \pm K)x}}$ are $\ket{\chi_{K} (z) e^{i(k \pm K)x}}$. We ignore the dependence of the confinement wavefunction $\chi$ on $k$. Further, we have used $\chi_{+ K} (z)=\chi_{- K} (z)$ because the non-perturbed structure is symmetric under $x$-reflection. 
The lossy modes  at $\ket{e^{ikx}}$ are modeled by $\ket{\chi^{(0)}(z) e^{ikx}}$, $\ket{\chi^{(1)}(z) e^{ikx}}$, \ldots. 
The matrix element scattering $\ket{\chi_K (z) e^{i(k\pm K)x}}$ into 
the $p$-th lossy mode $\ket{\chi^{(p)}(z) e^{ikx}}$ is given by 
$\braket{\chi^{(p)}(z)}{\chi_K (z)} \bra{1} u(x) \ket{e^{i(k\pm K)x}} = c^p U_{\pm 1}$. Notice that the first Fourier coefficients of $u(x)$ are related by $U_{-1}=U_{1}^\ast$.
For simplicity, in the following we consider the coupling to only the $0$-th lossy mode; the analysis can be extended to coupling to many lossy Fabry-Pérot modes in a straightforward way.

Restricted to the space spanned by three modes $\ket{+1} = \ket{\chi_K(z) e^{i(k+K)x}}$, $\ket{-1}= \ket{\chi_K({z}) e^{i(k-K)x}}$ and $\ket{0}= \ket{\chi^{0}(z) e^{ikx}}$, a general wavefunction can be written as
\begin{equation}
    |\tilde{\Phi} \rangle = \tilde{\phi}^k_{+1} \ket{+1} + \tilde{\phi}^k_{-1} \ket{-1} + \tilde{\phi}^k_{0} \ket{0}.
    \label{eq:threemode0}
\end{equation}
The evolution of the coefficients $(\tilde{\phi}^k_{+1},\tilde{\phi}^k_{-1},\tilde{\phi}^k_{0})$ follows the Schrödinger-like equation
\begin{equation}
   i \frac{d}{dt} 
    \begin{pmatrix}
        \tilde{\phi}^k_{+1} \\
        \tilde{\phi}^k_{-1} \\
        \tilde{\phi}^k_0
    \end{pmatrix}
    =
     \begin{pmatrix}
        \omega_{K}+vk & U_2 & c_0 U_1 \\
        U_2^{\ast} & \omega_{K} - vk & c_0 U_1^\ast \\
        c_0^\ast U_1^\ast & c_0^\ast U_1 & \omega_0 - i\gamma_0
    \end{pmatrix} 
    \begin{pmatrix}
        \tilde{\phi}^k_{+1} \\
        \tilde{\phi}^k_{-1} \\
        \tilde{\phi}^k_0
    \end{pmatrix},
    \label{eq:threemode}
\end{equation}  
where $\gamma_0$ is the decay rate of the low-momentum mode $\ket{\chi^{(0)} (z) e^{ikx}}$, which is assumed to vary negligibly for small $k$. We have also linearized the dispersal relation so that $\omega_{K+k} \approx \omega_K + v k$ and $\omega_{-K+k} \approx \omega_K - v k$, with $v$ being the light velocity near $K$. By adjusting a global phase, one can also assume that $\omega_K=0$ and $\omega_0$ can be then replaced by the difference in frequency $\Delta=\omega_0-\omega_K$.

Assuming that the decay rate of the lossy mode $\gamma_0$ is much faster than the matrix element $U_2$, one can adiabatically eliminate $\tilde{\phi}^k_0$. 
This is done by solving $\tilde{\phi}^k_0$ in terms of $\tilde{\phi}^k_{+1}$ and $\tilde{\phi}^k_{-1}$ as
\begin{equation}
    \tilde{\phi}^k_0 = -ic_0^\ast \int_{0}^{t} d \tau e^{-(\gamma_0 + i \Delta) \tau} \begin{pmatrix} U_1^* & U_1 \end{pmatrix} \begin{pmatrix} \tilde{\phi}_{+1}^k(t-\tau)  \\ \tilde{\phi}_{-1}^k (t-\tau) \end{pmatrix}.
\end{equation}
As the decaying of the lossy mode $\gamma_0$ is fast in comparison to the dynamics of the confined mode, one can make the Markovian approximation $\tilde{\phi}_{+1}^k(t-\tau) \approx \tilde{\phi}_{+1}^k(t)$,  $\tilde{\phi}_{-1}^k (t)=\tilde{\phi}_{+1}^k(t-\tau)$ and $\int_{0}^t d \tau e^{-(\gamma_0 + i \Delta) \tau} \approx \int_{0}^\infty d \tau e^{-(\gamma_0 + i \Delta) \tau} =1/(\gamma_0 + i \Delta) \approx 1/\gamma_0$. We have we also assumed $\gamma_0 \gg \Delta$ in the last approximation. In the end, we then obtain
\begin{equation}
    \tilde{\phi}^k_0 = \frac{-i c_0^\ast}{\gamma_0} \begin{pmatrix} U_1^* & U_1 \end{pmatrix} \begin{pmatrix} \tilde{\phi}_{+1}^k  \\ \tilde{\phi}_{-1}^k  \end{pmatrix}.
    \label{eq:markov}
\end{equation}
Inserting~\eqref{eq:markov} into~\eqref{eq:threemode}, we obtain the evolution equation for $\tilde{\phi}^k_{+1}$ and $\tilde{\phi}^k_{-1}$ in the form
\begin{equation}
\frac{d}{dt}     
    \begin{pmatrix}
        \tilde{\phi}_{+1}^k \\
        \tilde{\phi}_{-1}^k \\
    \end{pmatrix}
    =
    - i H
    \begin{pmatrix}
        \tilde{\phi}_{+1}^k \\
        \tilde{\phi}_{-1}^k \\
    \end{pmatrix}
\end{equation}
with the non-hermitian Hamiltonian in momentum space
\begin{equation}
    H= 
    \begin{pmatrix}
        +ivk & U_2  \\
        U_2^{\ast} & -ivk  \\
    \end{pmatrix} 
    -     i \gamma
    \begin{pmatrix}
        1 & e^{+i 2\varphi_1} \\
        e^{-i 2\varphi_1} & 1
    \end{pmatrix}
    \label{eq:Dirac-complex-U2}
\end{equation}
where $\varphi_1$ is defined by $U_1=  \abs{U_1}e^{i \varphi_1}$ and $\gamma= \abs{c_0}^2 \frac{\abs{U_1}^2}{\gamma_0}$. It is interesting to notice that the indirect loss rate $\gamma$ for the two modes $\ket{\chi_K(z) e^{\pm iKx}}$ is inversely proportional to the lossy rate $\gamma_0$ of the mode $\ket{\chi^{(0)}(z)e^{ikx}}$, indicating an analogy of the Zeno effect in quantum system~\cite{quantumZeno}. Indeed, a strong "measurement regime" that corresponds to a very leaky channel, $\gamma_0\gg \abs{U_1}$, will "freeze" the population of  $\ket{\chi_K(z) e^{\pm iKx}}$ because $\gamma\approx 0$. A similar setup, combining lossless waveguides and a lossy one, has been recently proposed to demonstrate the optical Zeno effect~\cite{photonicZeno}.

The Fourier coeffiecient $U_2$ in~\eqref{eq:Dirac-complex-U2} is generally a complex number. We denote $U_2=\abs{U_2} e^{i \varphi_2}$ and eliminate the phase $\varphi_2$ by the following unitary transformation,
\begin{equation}
    \begin{pmatrix}
        \tilde{\phi}_+^k \\
        \tilde{\phi}_-^k
    \end{pmatrix}
    =
    \begin{pmatrix}
        e^{+i \varphi_2/2} & 0\\ 
        0 & e^{-i \varphi_2/2}
    \end{pmatrix}
    \begin{pmatrix}
        {\phi}_+^k \\
        {\phi}_-^k
    \end{pmatrix}.
\end{equation}
One then obtains the non-hermitian Hamiltonian for $\phi_{+1}^k$ and $\phi_{-1}^k$ as

    \begin{equation}
    H= 
    \begin{pmatrix}
        +ivk & \abs{U_2}  \\
        \abs{U_2} & -ivk  \\
    \end{pmatrix} 
    -     i \gamma
    \begin{pmatrix}
        1 & e^{+i \varphi} \\
        e^{-i \varphi} & 1
    \end{pmatrix}
\end{equation}
with $\varphi= 2 \varphi_1-\varphi_2$. This is the effective Hamiltonian~\eqref{eq.Hpho} introduced in the main text with $\kappa=\abs{U_2}$.

\subsection{Derivation of the farfield and nearfield intensity from the microscopic description}
\label{sec:farfield-microscopic}
We start with remarking again that the loss due to radiation into the farfield of the spinor polaritons $\Psi$ inherits directly from the loss of the photonic component $\Phi$. Therefore the farfield pattern of the polariton can be understood directly from its photonic components.

Generally, the effective wave function $\Phi(x)$ is a superposition of different wavevectors $k$,
\begin{equation}
\Phi(x) = \sum_{k} e^{ikx} 
\begin{pmatrix} 
\phi_{+1}^k \\ \phi_{-1}^k
\end{pmatrix}.
\label{eq:fourier}
\end{equation}
At the microscopic level, this is a plane wave of
\begin{equation}
    \ket{\Phi} = \sum_{k} \phi^k_{+1} e^{i \varphi_2/2} \ket{e^{i(k+K)x}} + \phi^k_{-1} e^{-i \varphi_2/2} \ket{e^{i(k-K)x}},
\end{equation}
if we ignore the confining mode function in the $z$-direction and  the lossy mode  as comparison to~\eqref{eq:threemode0}. The latter is only relevant to the lossy dynamics.
Explicitly in terms of the electric field, one has
\begin{equation}
\vec{E}^{\text{near}} \propto  \left(\phi_{+1}^k  e^{ i \varphi_2/2} e^{ iKx}  + \phi_{-1}^k  e^{-i \varphi_2/2} e^{-iKx}\right) e^{ikx} \vec{u}_y,
\end{equation}
where $\vec{u}_y$ denotes the polarization direction.
Or using the real space representation $\Phi(x)$, we can write
\begin{equation}
\vec{{E}}^{\text{near}} \propto   \left[\phi_{+1} (x) e^{i \varphi_2/2} e^{iKx} + \phi_{-1} (x)  e^{-i \varphi_2/2} e^{-iKx}\right] \vec{u}_y.  
\label{eq:evelope-core}
\end{equation}
Recall that we are working in the regime where $\phi^k_{+1}$ and $\phi^k_{-1}$ are only significant at wavelengths $k \ll K$.  Equivalently, $\phi_{+1}(x)$ and $\phi_{-1}(x)$, which are referred to as spatial \emph{envelope functions}, vary much slower than $e^{\pm i Kx}$, which are referred to as \emph{core functions}.
Averaging out the fast fluctuation at the wavevector $K$, the nearfield intensity can then be obtained as:
\begin{equation} \label{eq.nearfield}
    I^{\text{near}} (x) = \abs{\phi_{+1} (x)}^2 + \abs{\phi_{-1} (x)}^2.
\end{equation}
This agrees with the formal derivation from the effective photonic Dirac equation~\eqref{eq:continuity}.

As for the farfield, we notice that the farfield is the observation of the lossy mode $\phi^k_0$, which is given by~\eqref{eq:markov} for a single $k$. With a superposition of different wavevectors~\eqref{eq:fourier}, one has
\begin{equation}
    \phi_0(x) \propto \sum_{k} [U_1^\ast e^{+i \varphi_2/2} \phi^k_{+1} + U_1 e^{-i \varphi_2/2} \phi^k_{-1}].
\end{equation}
Explicitly in terms of the electric field, this corresponds to 
\begin{equation}
    \vec{E}^{\text{far}} (x) \propto \sum_k e^{ikx} (\phi^k_{+1} + e^{i \varphi} \phi^k_{-1}) \vec{u}_y,
\end{equation}
where we have again used $\varphi=2\varphi_1-\varphi_2$.
The farfield intensity is then obtained as
\begin{equation} \label{eq.farfield}
    I^{\text{far}} (x) \propto \abs{\phi_{+1}(x) + e^{i \varphi} \phi_{-1}(x)}^2,
\end{equation}
which coincides with~\eqref{eq:farfield1}.

\subsection{Design and numerical simulations for pratical realizations}\label{sec:design}
We propose realistic a stack of dielectric layers and quantum wells, based on the experimental works from~\cite{Ardizzone_Nature2022,gianfrate2023optically}. The sample stack is composed of a waveguide core made of 12 GaAs quantum wells (QWs) 20 nm thick and 13 Al$_{0.4}$Ga$_{0.6}$As barriers 20 nm thick grown on an Al$_{0.8}$Ga$_{0.2}$As 500 nm thick cladding layer. The cladding and the GaAs substrated are seperated by a 50 nm-AlAs layer.  The whole stack is capped by a 10 nm GaAs layer. 

For numerical simulations of photonic modes in the gratings, the excitonic resonances in the GaAs QWs are removed in the dielectric function. The photonic modes are calculated by numerical simulations based on Rigorous Coupled-Wave Analysis (RCWA) method  with the S\textsuperscript{4} package provided by the Fan Group at the Stanford Electrical Engineering Department~\cite{Liu2012}. The refractive index of the QWs, barriers and the cladding are: $n_\text{QWs}=3.547 + 0.0001i$, $n_\text{barrier} = 3.3$, $n_\text{cladding} = 3.063$. The imaginary part in the refractive index of the QWs are simply added to probe the photonic modes in absorption simulations. We only calculate TE (transverse electric) photonic modes since the TM (transverse magnetic) photonic modes are inefficiently coupled to in-plane excitonic dipoles of the QWs. 

We also employ RCWA method for the numerical simulations of polariton modes. To do so, a  Lorentz oscillator at 1527.4 meV with 0.001 eV$^2$ oscillator strength and 0.35 meV linewidth is added in the dielectric function of the QWs. This excitonic resonance corresponds to the heavy-hole excitons of the QWs. 
\begin{figure*}[ht!]
\centering
\includegraphics[width=0.8\linewidth]{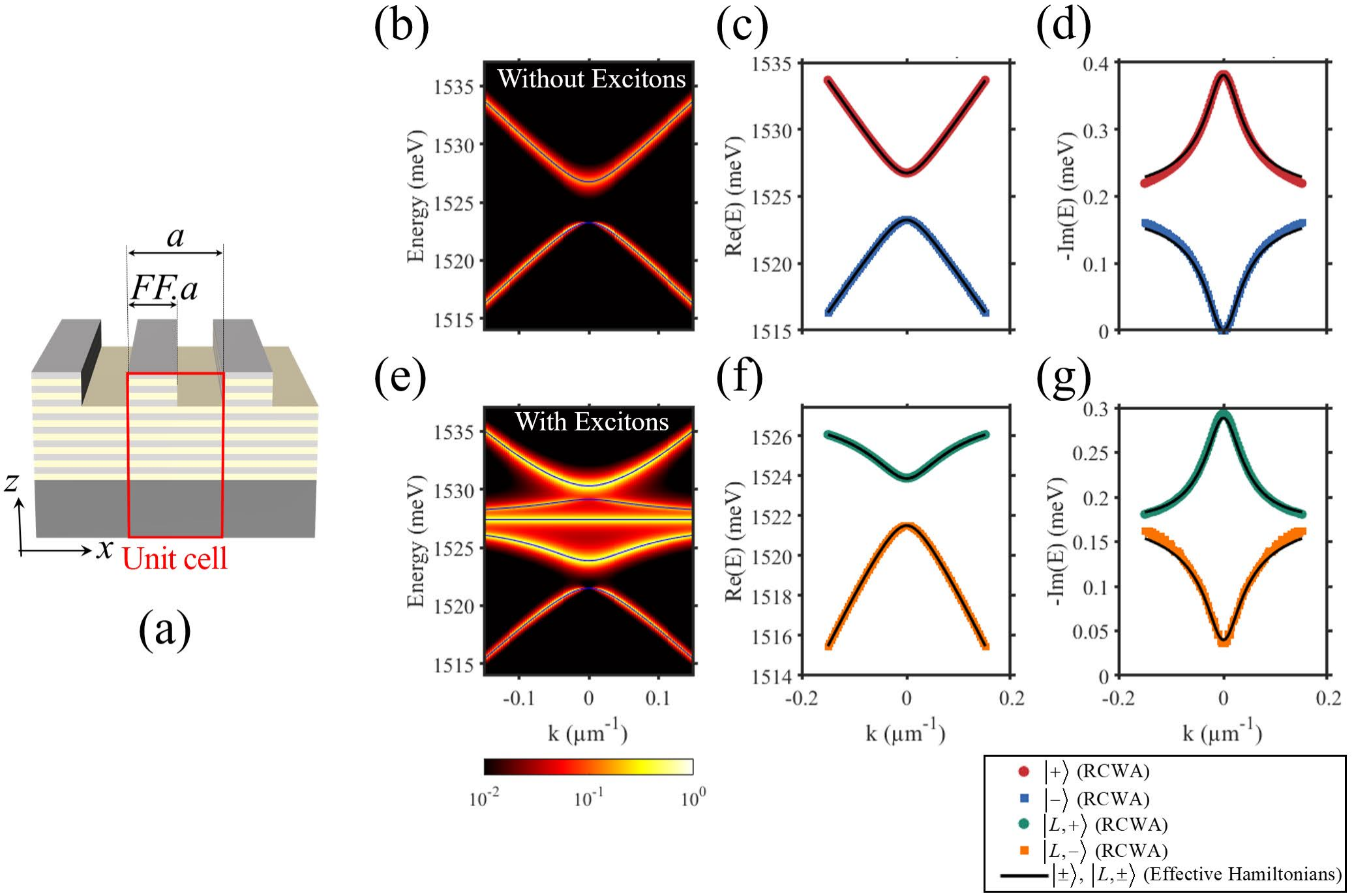}
\caption{\textbf{RCWA simulations versus effective Dirac-photon theory.} (a) Sketch of the grating design. (b,e) Photonic and polaritonic dispersions, numerically calculated by RCWA, for $a=243$ nm, $FF=0.34$, and 110 nm of etching. The solid blue lines correspond to theoretical dispersion from the effective theory. (c,d) Real and imaginary part for the energy of the photonic modes. Symbols are numerical results extracted from RCWA simulation in (b). Solid black line is the theoretical prediction using~\eqref{eq.Hpho}.   (f,g) Real and imaginary part for the energy of the two lower polaritonic modes. Symbols are numerical results extracted from RCWA simulation in (e). Solid black line is the theoretical prediction using  \eqref{eq.HpolDirac}. The parameters for effective theory are $\hbar \gamma = 0.19$ meV, $\hbar \kappa = 1.75$ meV, $\hbar v = 56.61$ meV $\mu$m, $\hbar \Omega = 3.2$ meV, $\hbar \omega^{(0)_X} = 2.41$ meV,  $\hbar \gamma_{nr} = 0.18$ meV and $\varphi=0$.}
\label{fig_SimulVStheory}
\end{figure*}
\begin{figure*}[ht!]
\centering
\includegraphics[width=0.8\linewidth]{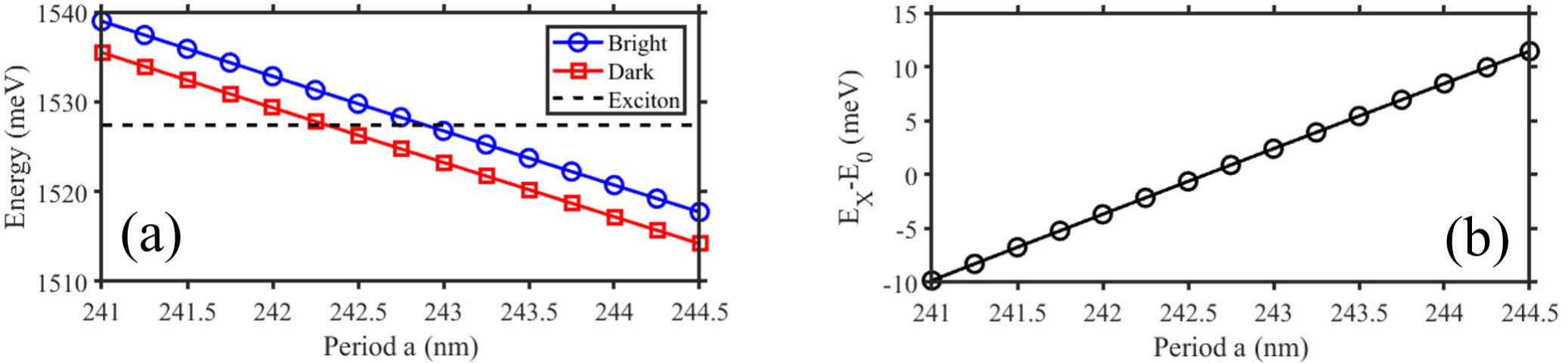}
\caption{\textbf{Varying the exciton-photon energy detuning.} (a) Energy at $k=0$ of the photonic modes, given by RCWA simulation results as function of the period $a$. (b) The dependence of the exciton-photon detuning as function of $a$. The RCWA are performed for grating of filling fraction $FF=0.37$ and the etching depth of 110 nm.}
\label{fig_scanninga}
\end{figure*}
\begin{figure*}[ht!]
\centering
\includegraphics[width=0.8\linewidth]{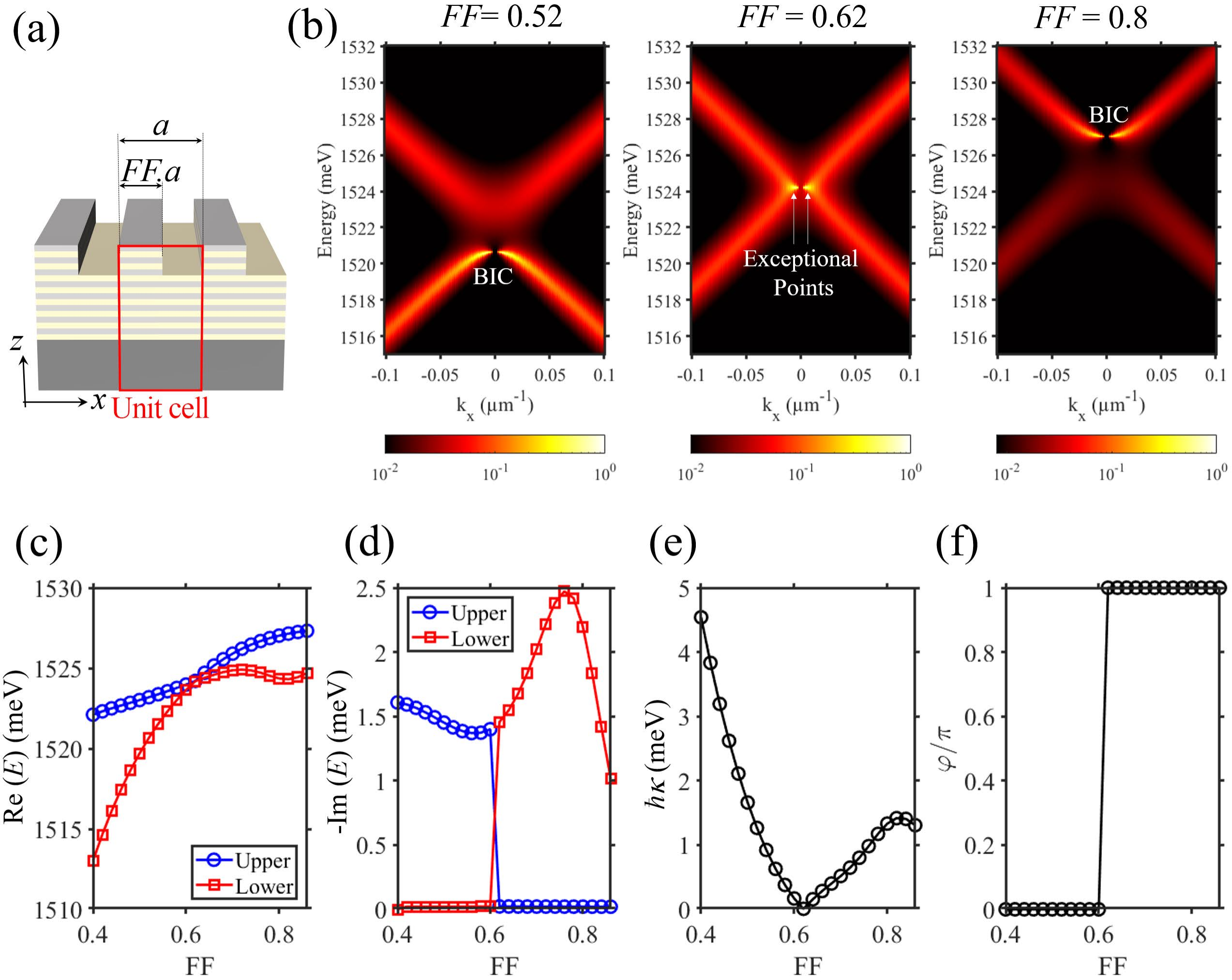}
\caption{\textbf{Continuously tuning  $\kappa$ and implementing $\pi$-jump to $\varphi$ via the filling fraction $FF$.} (a) Sketch of the grating design. (b) Photonic dispersion, numerically calculated by RCWA, for $FF=0.52$, $FF=0.62$ and $FF=0.8$. (c,d) Real and imaginary part for the energy of the photonic modes when scanning $FF$ from 0.4 to 0.86. (e,f)  $\kappa$ and $FF$ when  scanning $FF$ from 0.4 to 0.86.}
\label{fig_scanningFF}
\end{figure*}
\begin{figure*}[ht!]
\centering
\includegraphics[width=0.8\linewidth]{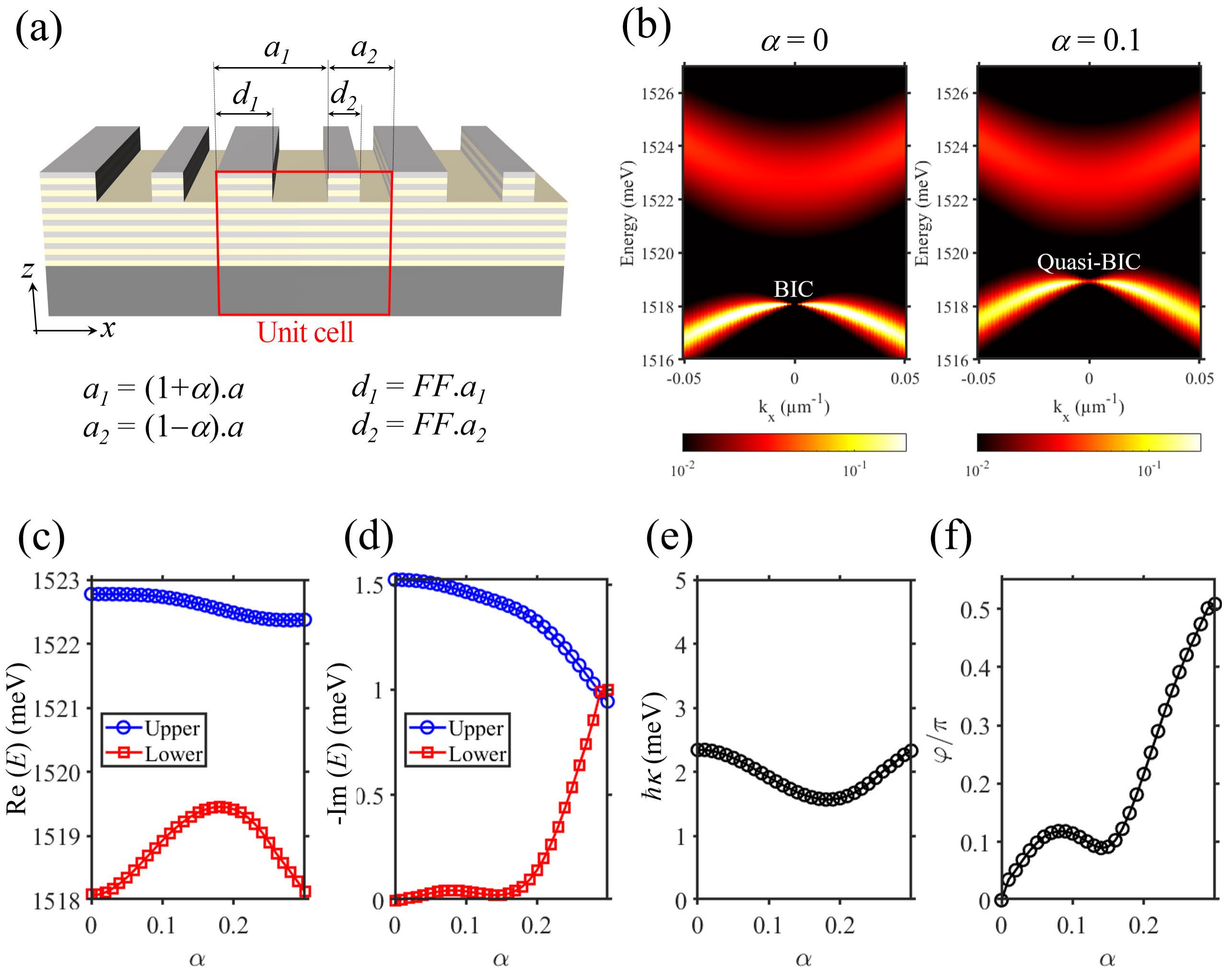}
\caption{\textbf{Continuously tuning $\varphi$ via the symmetry-breaking coefficient $\alpha$.} (a) Sketch of the grating design with double-period symmetry-breaking. b) Photonic dispersion, numerically calculated by RCWA, for $\alpha=0$ and $\alpha=0.1$. (c,d) Real and imaginary part for the energy of the photonic modes when scanning $\alpha$ from 0 to 0.3. (e,f)  $\kappa$ and $FF$ when  scanning $FF$ from 0 to 0.3.}
\label{fig_scanning_alpha}
\end{figure*}

\subsection{Numerical RCWA simulation versus developed effective theory}\label{sec:simulVStheory}
To validate the effective Dirac theory for the guided photonic modes~\eqref{eq.Hpho} and the polaritonic modes~\eqref{eq.HpolDirac}, we perform RCWA simulations of the grating of period $a=243$ nm, filling fraction $FF=0.37$ and 110 nm of etching depth. The numerical results of photonic and polaritonic modes, together with the fittings using the effective Hamiltonians are presented in Fig.~\ref{fig_SimulVStheory}. It shows that both simulated photonic and polaritonic modes are perfectly followed by our analytical model with:  $\hbar \gamma = 0.19$ meV, $\hbar \kappa = 1.75$ meV, $\hbar v = 56.61$ meV $\mu$m, $\hbar \Omega = 3.2$ meV, $\hbar \omega^{(0)_X} = 2.41$ meV,  $\hbar \gamma_{nr} = 0.18$ meV and $\varphi=0$. Importantly, the vanishing of the intensity in the lower antisymmetric polariton branch $\ket{L,-}$ at $k=0$ in Fig.~\ref{fig_SimulVStheory}(c) confirms the infinite radiative lifetime of these polaritons that are inherited from the photonic BIC. Therefore the finite linewidth of $\ket{L,-}$ at $k=0$ [see Fig.~\ref{fig_SimulVStheory}(g)] is purely nonradiative and is inherited by the excitonic component. These results are in good agreement with the experimental observations in \cite{Dang_AdvOptMat2022}.

\subsubsection{Varying the exciton-photon energy detuning}\label{sec:scanning_a}
By scanning the period $a$ of the same design (i.e. $FF$, etching depth), the energy detuning $\hbar \omega^{(0)}_X$ between the exciton energy $E_X$ and the mid gap of photonic modes $E_0=\text{Re}{\left(\hbar\omega_+ + \hbar\omega_-\right)/2}$ can be freely varied, with all other parameters unchanged. This is clearly evidenced in Fig.~\ref{fig_scanninga} that reports the RCWA results of the energy of $\ket{\pm}$ symmetric and antisymmetric photonic modes at $k=0$ when scanning the period $a$ from 241 nm to 244.5 nm.  

\subsubsection{Tuning the diffractive coupling  $\kappa$ and implementing $\pi$-jump to $\varphi$}\label{sec:scanningFF}
As previously reported~\cite{Lu_PhotRes2020,Ardizzone_Nature2022}, the value of  $\kappa$ can be continuously tuned by modifying the filling fraction $FF$. To illustrate this effect in the case of our design, we keep the period $a=250$ nm  fixed and monitoring the modification of the photonic modes when scanning $FF$ from 0.4 to 0.86. This scanning induces a band-inversion to the photonic modes~\cite{Lu_PhotRes2020}. Indeed, simulated photonic dispersions with $FF=$0.52, 0.62 and 0.8 are shown in Fig.~\ref{fig_scanningFF}(b). These results represent respectively three case: i) photonic BIC in the lower band, corresponding to non-zero  $\kappa$ and $\varphi=0$; ii) gap closing and formation of Exceptional Points~\cite{Lu_PhotRes2020}, corresponding to  $\kappa=0$; iii) photonic BIC in the upper band, corresponding to non-zero  $\kappa$ and $\varphi=\pi$. The values of  $\kappa$ and $\varphi$ as the function of $FF$ are extracted from the real-part [see Fig.~\ref{fig_scanningFF}(c)] and imaginary-part [see Fig.\ref{fig_scanningFF}(d)] of the photonic modes. These results, presented in  Figs.~\ref{fig_scanningFF}(e) and~\ref{fig_scanningFF}(f), show that $\kappa$ is continuously tuned between 0 and 5 meV; and $\varphi$ undergoes a $\pi$ jump when  $\kappa=0$ ($FF$=0.62).

\subsubsection{Tuning continuously the phase $\varphi$}\label{sec:scanningalpha}
To obtain $\varphi$ that is not a multiple of $\pi$, we break the in-plan mirror symmetry $-x\rightarrow x$ of the grating. This is achieved by employing double-period the design~\cite{Nguyen2018}: each unitcell now consists of two sub-cell of period $(1+\alpha)a$ and $(1-\alpha)a$ with $\alpha$ being the symmetry-breaking coefficient [see Fig.~\ref{fig_scanning_alpha}(a)]. Consequently, the phase-shift $\varphi$ can be continuously tuned by changing $\alpha$. 

To illustrate this effect, we keep $a=250$ nm, $FF=0.45$, and monitoring the modification of the photonic modes when scanning $\alpha$ from 0 to 0.25. As expected, a non-zero $\alpha$ turn a BIC into quasi-BIC [see Fig.~\ref{fig_scanning_alpha}(b)]:  the farfield at $k=0$ of the quasi-BIC is not-vanished and the photonic band exhibits non-zero linewidth.  The values of  $\kappa$ and $\varphi$ as the function of $\alpha$ are extracted from the real-part [see Fig.~\ref{fig_scanning_alpha}(c)] and imaginary-part [see Fig.~\ref{fig_scanning_alpha}(d)] of the photonic modes. These results, presented in  Figs.~\ref{fig_scanning_alpha}(e) and~\ref{fig_scanning_alpha}(f), show that while  $\kappa$ only undergoes small variation, $\varphi$ is continuously tuned between 0 and $\pi/2$. Interestingly, for $\ket{L,-}$ is still almost lossless, thus being quasi-BIC, until $\alpha$ exceeds 0.15 [see Fig.~\ref{fig_scanning_alpha}(d)], that corresponds to $\varphi< 0.15\pi$ [see Fig.~\ref{fig_scanning_alpha}(f)]. 

\subsection{Numerical mean field methods} \label{sec.numMF}
Equations~\eqref{eq:GP1} and~\eqref{eq:GP2} form a coupled nonlinear systems of equation to be solved for $n_R$ and $\Psi$. Numerical simulations are performed using fast Fourier transform spectral methods in space and an explicit Runge-Kutta 4th-5th order formula, the Dormand-Prince pair, in time. We apply damped boundary conditions in real and reciprocal space to avoid periodic boundary effects coming from the spectral methods, and always start from random white noise initial conditions. The gridpoint spacing $\Delta x$ is chosen small enough to encompass the necessary features in momentum space. In our case, $\Delta x = 4$ $\mu$m was sufficient. A variable timestep is set by the MATLAB\textregistered~\texttt{ode45} solver to reach the desired accuracy.

\section{Author statements} 

\begin{acknowledgement}
We acknowledge fruitful discussions with Davide Nigro and Dario Gerace. 
\end{acknowledgement}

\begin{funding}
H.S. acknowledges the project No. 2022/45/P/ST3/00467 co-funded by the Polish National Science Centre and the European Union Framework Programme for Research and Innovation Horizon 2020 under the Marie Skłodowska-Curie grant agreement No. 945339.
H.C.N. acknowledges 
the Deutsche Forschungsgemeinschaft (DFG, German Research Foundation, project numbers 447948357 and 440958198), 
the Sino-German Center for Research Promotion (Project M-0294), 
the German Ministry of Education and Research (Project QuKuK, BMBF Grant No. 16KIS1618K) 
and the ERC (Consolidator Grant 683107/TempoQ).
\end{funding}

\begin{authorcontributions}
H.S., H.C.N. and H.S.N. contributed equally to the developed theoretical description and writing of the manuscript. All authors have accepted responsibility for the entire content of this manuscript and approved its submission.
\end{authorcontributions}

\begin{conflictofinterest}
Authors state no conflict of interest.
\end{conflictofinterest}

\begin{dataavailabilitystatement}
The datasets generated during and/or analyzed during the current study are available from the corresponding author on reasonable request.
\end{dataavailabilitystatement}

\bibliographystyle{ieeetr}

\end{document}